\documentclass[aps,preprint]{revtex4}%
\usepackage{amsfonts}
\usepackage{amsmath}
\usepackage{amssymb}
\usepackage{graphicx}
\def\be{\begin{eqnarray}}
\def\ee{\end{eqnarray}}
\renewcommand{\v}[1]{{\bf #1}}

\newcommand{\s}{{\sigma}}

\newcommand{\nn}{\nonumber\\}
\newcommand{\Eq}[1]{Eq.~(\ref{#1})}

\newcommand{\p}{\partial}

\newcommand{\ra}{\rightarrow}

\newcommand{\cM}{ {\cal M} }

\newcommand{\cP}{ {\cal P} }

\newtheorem{theorem}{Theorem}

\newtheorem{example}[theorem]{Example}

\newenvironment{proof}[1][Proof]{\noindent\textbf{#1.} }{\ \rule{0.5em}{0.5em}}

\begin{document}

\title{Eigen Wavefunctions of a Charged Particle Moving in a Self-Linking
Magnetic Field}
\author{Dah-Wei Chiou}
 \email{dwchiou@socrates.berkeley.edu}
 \affiliation{Department of Physics\\ University of California at Berkeley\\ Berkeley, CA 94720 USA}
\author{Wu-Yi Hsiang}
 \email{hsiang@math.berkeley.edu}
 \affiliation{Department of Mathematics\\ University of California
 at Berkeley\\ Berkeley, CA 94720 USA}
\author{Dung-Hai Lee}
 \email{dunghai@socrates.berkeley.edu}
 \affiliation{Department of Physics\\ University of California at Berkeley\\ Berkeley, CA 94720 USA}


\begin{abstract}
In this paper we solve the one-particle Schr\"{o}dinger equation
in a magnetic field whose flux lines exhibit mutual linking. To
make this problem analytically tractable, we consider a
high-symmetry situation where the particle moves in a three-sphere
$(S^3)$. The vector potential is obtained from the Berry
connection of the two by two Hamiltonian $H(\v r)=\hat{h}(\v r)
\cdot\vec{\sigma}$, where $\v r\in S^3$, $\hat{h}\in S^2$ and
$\vec{\sigma}$ are the Pauli matrices. In order to produce linking
flux lines, the map $\hat{h}:S^3\rightarrow S^2$  is made to
possess nontrivial homotopy. The problem is exactly solvable for a
particular mapping ($\hat{h}$) . The resulting eigenfunctions are
$SO(4)$ spherical harmonics, the same as those when the magnetic
field is absent. The highly nontrivial magnetic field lifts the
degeneracy in the energy spectrum in a way reminiscent of the
Zeeman effect.

\end{abstract}
\maketitle
\tableofcontents

\section{Introduction}

It is remarkable that the basis of some of the most fundamental
theories of nature, such as general relativity and gauge theories,
is the mathematics of fiber bundles. A fiber bundle is locally
characterized by its curvature tensor (the strength of the gauge
field), and globally by its topological invariants.

For example the quantization of the magnetic charge of a Dirac
monopole \cite{Dirac:1931, Wu-Yang:1975} is equivalent to the
quantization of the first Chern number,
$\int_{S^2}ch_{1}(\mathcal{F})$, for the $U(1)$ bundle over $S^2$
(See \cite{Nakahara} and Subsection \ref{subsec: Chern number}
below).
Another well-known example is the $SU(2)$ instanton
\cite{Belavin:1975}. The instanton number is given by the integral
of the second Chern character, $\int_{S^{4}%
}ch_{2}(\mathcal{F})$, for the $SU(2)$ bundle over $S^{4}$
\cite{Nakahara}.

Under the influence of such topologically nontrivial background
gauge field, the dynamics of the particles that interact with it
is profoundly affected.  For example Wu and Yang have shown that
in the presence of a Dirac monopole the orbital angular momentum
quantum number of a charge $e$ particle can become odd integer
multiples of $\hbar$ \cite{Wu-Yang:1975}. Under such condition we
need to generalize ``wave function'' to ``\emph{wave section}''
which is patch-wise defined in space \cite{Wu-Yang:1976}. The
eigensections of the angular momentum are called \emph{monopole
harmonics} $Y_{q,l,m}$, where $q=e\times$the magnetic charge. The
allowed values of $q$ are multiples of half-integers, and the
allowed values of $l$ and $m$ are: $l=\left\vert q\right\vert
,\,\left\vert q\right\vert +1,\,\left\vert q\right\vert +2,\,...$
and $m=-l,\,-l+1,\,...,\,l$. Thus half-integer $q$ gives
half-integer $l$, which means $Y_{q,l,m}$ can form a complete
representation of $SU(2)$ whereas ordinary spherical harmonics
$Y_{l,m}$ can only represent $SO(3)$. Therefore, without
introducing the spinor structure, applying nontrivial field
strength alone can lift the wave function representation to
fulfill the universal covering group.

As stated above, the first Chern number counts the number of
fundamental monopoles enclosed by a 2-dimensional closed surface.
In a 3-dimensional space, field configurations with the same Chern
invariant may have different linking topologies for the loops of
the magnetic lines. \emph{Chern-Simons invariant} further
distinguishes this topology. In this paper, in the spirit of Wu
and Yang, we investigate the effect of self-linking magnetic field
on charged particle dynamics. To make this problem tractable
analytically, we consider a high- symmetry situation where the
particle moves in a three-sphere $(S^3)$. The vector potential is
obtained from the Berry connection of the two by two Hamiltonian
$H(\v r)=\hat{h}(\v r) \cdot\vec{\sigma}$, where $\v r\in S^3$,
$\hat{h}\in S^2$ and $\vec{\sigma}$ are the Pauli matrices
\cite{HsiangLee}. In order to produce linking flux lines, the map
$\hat{h}:S^3\rightarrow S^2$  is made to possess nontrivial
homotopy. For a particular mapping ($\hat{h}$) the problem is
exactly solvable. The resulting eigenfunctions are $SO(4)$
spherical harmonics, the same as those when the magnetic field is
absent. The highly nontrivial magnetic field lifts the degeneracy
in the energy spectrum in a way reminiscent of the Zeeman effect.

This paper is organized as follows. In Section \ref{sec: Berry's
phase}, we briefly review the physics of Berry's phase.  In
Section \ref{sec: Chern-Simons}, we study the topological
classifications by Chern and Chern-Simons invariants. Later, in
Section \ref{sec: Shrodinger}, with the help of Berry connection,
we construct the Schr\"{o}dinger equation in the presence of
magnetic field with nontrivial Chern-Simons invariant. We solve
the Schr\"{o}dinger equation without magnetic field in Section
\ref{sec: spherical harmonics} and get $SO(4)$ spherical
harmonics. Finally, in Section \ref{sec: nonzero field}, the
Schr\"{o}dinger equation with magnetic field of nonzero
Chern-Simons invariant is solved. We conclude in Section \ref{sec:
discussion} with a discussion of our results.

\section{Berry's Phase and its Geometric
Nature\label{sec: Berry's phase}}

In quantum mechanics, the overall phase of a wave function is
often regarded as an irrelevant factor. Berry pointed out that the
phase may have observable consequences if the system undergoes an
adiabatic change \cite{Berry:1984}. In fact, Berry's phase has a
deep geometrical meaning with gauge nature \cite{Simon:1983} and
can be used to give the electromagnetic potential as showed in
Example \ref{ex: monopole 1}.

\subsection{Geometric (Berry's) Phase}

Consider a Hamiltonian $H(\v x)$ which depends on an external set
of parameters
$\v x=(x_{1},\ldots,x_{n})$. If $\v x(t)$ changes \textquotedblleft adiabatically\textquotedblright%
\ (slowly)\ along a closed path $C:\v x(t)\ \ t\in\lbrack
t_{i},t_{f}]$, with $\v x(t_{i})=\v x(t_{f})$ in the parameter
space, then the solution of the equation $i\partial
_{t}|\psi(t)\rangle=H(\v x(t)) |\psi(t)\rangle$ is
\begin{align}
|\psi(t_f)\rangle &  = e^{-i\int_{t_i}^{t_f} dtE(\v
x(t))}e^{i\gamma_{C}}|\psi(t_i)\rangle,
\end{align}
where \be\gamma_{C}  &  =\oint_{C}A_{\mu}(x)dx^{\mu},{\rm~~
with~~}
A_{\mu}(x)=-i\langle\psi(x)|\partial_{\mu}\psi(x)\rangle.\label{bp}\ee
In \Eq{bp},  $|\psi(\v x)\rangle$ is a \emph{normalized}%
\textit{, }\emph{non-degenerate} eigenstate satisfying
\begin{equation}
H(\v x)|\psi\rangle=E(\v x)|\psi(\v x)\rangle.
\end{equation}
In addition, the phase of $|\psi(\v x)\rangle$ is chosen so that
$\partial_{\mu}|\psi(\v x)\rangle$ is defined for $\v x\in C$.

Therefore, after $x(t)$ completes a closed circuit, the eigenstate
$|\psi\rangle$ returns to its initial value with an additional
phase, of which $e^{-i\int dtE}$ is due to the time revolution
while the extra phase $e^{i\gamma_{C}}$ is called \emph{geometric
(Berry's) phase}.

\subsection{The Berry Connection and Berry Curvature; the Abelian Case}

If we consider a local (gauge) transformation, $|\psi(\v
x)\rangle\rightarrow e^{i\theta(\v x)}|\psi(\v x)\rangle$, and use
$e^{i\theta(\v x)}|\psi(\v x)\rangle$ to compute $A_{\mu}$, we get
\begin{equation}
A_{\mu}(\v x)\rightarrow A_{\mu}(\v x)+\partial_{\mu}\theta(\v x),
\end{equation}
which shows that $A_{\mu}(x)$ transforms like a gauge field. In
the literature it is often called \emph{Berry connection}. By
drawing analogy with electromagnetism we can define the
\emph{Berry curvature} as :
\begin{equation}
F_{\mu\nu}=\partial_{\mu}A_{\nu}-\partial_{\nu}A_{\mu}.
\end{equation}
The Berry curvature is gauge-invariant and its integral over the
surface bounded by the closed path $C$ gives the Berry phase:
\begin{equation}
\gamma_{C}=\int_{surface}F_{\mu\nu}dx^{\mu}dx^{\nu}.
\end{equation}

\begin{example}
\label{ex: monopole 1} Consider the following $2\times2$
Hamiltonian:
\begin{equation}
H(x,y,z)=\vec{r}\cdot\vec{\sigma}=x\sigma_{x}+y\sigma_{y}+z\sigma_{z}=\left(
\begin{array}
[c]{cc}%
z & x-iy\\
x+iy & -z
\end{array}
\right)  ,
\end{equation}
where the $\sigma$'s are Pauli matrices. $H$ has two eigenvalues
$\pm \sqrt{x^{2}+y^{2}+z^{2}}\equiv\pm r$ and two eigenvectors
\begin{equation}
|\psi_{\pm}(x,y,z)\rangle=\frac{1}{N_{\pm}}\left(
\begin{array}
[c]{c}%
\frac{z\pm r}{x+iy}\\
1
\end{array}
\right)  . \label{spinor}
\end{equation}

Taking $|\psi_{\pm}(x,y,z)\rangle$ to compute
$A_{\mu}^{\pm}=-i\langle
\psi_{\pm}|\partial_{\mu}\psi_{\pm}\rangle$, we obtain
\begin{equation}
A_{x}=\frac{-y}{2r(r\mp z)},\quad A_{y}=\frac{x}{2r(r\mp z)},\quad
A_{z}=0,
\label{eq: monopole A}%
\end{equation}
from which
$F_{\mu\nu}=\partial_{\mu}A_{\nu}-\partial_{\nu}A_{\mu}$ is
obtained:
\begin{equation}
F_{xy}=\mp\frac{z}{2r^{3}},\quad F_{zx}=\mp\frac{y}{2r^{3}},\quad
F_{yz}=\mp\frac {x}{2r^{3}}.
\end{equation}
If we interpret $x$, $y$, $z$ as the spatial space coordinates,
then $\varepsilon^{\mu\nu\lambda}F_{\nu\lambda}$ is precisely the
magnetic field produced by a monopole at $r=0$.

\emph{Remark}: The connection $A_{\mu}$ in \Eq{eq: monopole
A} is singular in the north (south) pole direction ($z=\pm r$). In
order to derive a non-singular $A_{\mu}$, we can use
\Eq{spinor} in the southern (northern) hemisphere, and \be
|\psi_{\pm}(x,y,z)\rangle=\frac{1}{N_{\pm}}\left(
\begin{array}
[c]{c}%
1\\{x+iy\over r \pm z}
\end{array}
\right)   \label{spinor1} \ee in the northen (southern)
hemisphere. The vector potential derived from \Eq{spinor} and
\Eq{spinor1} differs by a gauge transformation along the equator
\cite{Wu-Yang:1975}. (Also see Example \ref{ex: monopole 2}
below.)


\end{example}

\subsection{The Berry Connection and Berry Curvature; the Non-Abelian Case}

This can be extended to non-Abelian case if we consider
$n$-fold degenerate eigenstates instead of a non-degenerate one.
Let  $|\psi^{\alpha}\rangle, ~\alpha=1,...,n$ represent the
members of the degenerate n-tuplet. The Berry connection is an
$n\times n$ matrix:
\begin{equation}
{(A_{\mu})^{\beta}}_{\alpha}=-i\langle\psi_{\alpha}|\partial_{\mu}\psi^{\beta
}\rangle.
\end{equation}

Under a local (gauge) unitary transformation in the internal space
(eigenspace),
$|\psi^{\prime\alpha}(x)\rangle={S(x)^{\alpha}}_{\beta}|\psi^{\beta
}(x)\rangle$.  The non-Abelian connection then transforms as:
\begin{align}
{(A_{\mu}^{\prime})^{\beta}}_{\alpha}  &  =-i\langle{S_{\alpha}}^{\gamma}%
\psi_{\gamma}|\partial_{\mu}({S^{\beta}}_{\delta}\psi^{\delta})\rangle
\nonumber\\
&  ={S^{\beta}}_{\delta}{(A_{\mu})}_{\gamma}^{\delta}{{S_{\alpha}}^{\gamma}%
}^{\ast}-i(\partial_{\mu}{S^{\beta}}_{\delta}){(A_{\mu})^{\delta}}_{\gamma
}{{S_{\alpha}}^{\gamma}}^{\ast}.
\end{align}
With
${{S_{\alpha}}^{\gamma}}^{\ast}={{S^{\dagger}}^{\gamma}}_{\alpha}$,
this gives
\be
{(A_{\mu}^{\prime})^{\beta}}_{\alpha}  &
={S^{\beta}}_{\delta}{(A_{\mu
})^{\delta}}_{\gamma}{(S^{\dagger})^{\gamma}}_{\alpha}-i(\partial_{\mu
}{S^{\beta}}_{\delta}){(A_{\mu})^{\delta}}_{\gamma}{(S^{\dagger})^{\gamma}%
}_{\alpha}
\ee
or
\be
A_{\mu}^{\prime}  &
=S\,A_{\mu}\,S^{\dagger}-i(\partial_{\mu
}S)A_{\mu}S^{\dagger}.\label{transform}
\ee
The non-Abelian
curvature is defined as
\begin{equation}
F_{\mu\nu}=\partial_{\mu}A_{\nu}-\partial_{\nu}A_{\mu}-i[A_{\mu},\,A_{\nu}].
\end{equation}
Using \Eq{transform}, it is simple to show that $F_{\mu\nu}$ transforms
covariantly under the gauge transformation; i.e.,%
\begin{equation}
F_{\mu\nu}^{\prime}=S\,F_{\mu\nu}S^{\dagger}.
\end{equation}

\begin{example}
\label{ex: gr 2}

The close connection between non-Abelian gauge theories and
general relativity has been noticed and elucidated by Yang
\cite{YM-GR-yang}. In particular, the Christoffel connection can be
interpreted as the non-Abelian gauge potential and the
Riemann-Christoffel curvature is proportional to the corresponding
field strength \cite{YM-GR-book}.

In this example we show that for a given (curved) d-dimensional space
there exists a $(d+1)\times (d+1)$ matrix Hamiltonian whose
Berry's connection is the Christoffel connection of the subject
space and the Riemann-Christoffel curvature is the minus Berry
curvature. This result allows us to further tighten the connection
between geometric phase and differential geometry.

Consider a d-dimensional (curved) space $\cM$ as being embedded in
a d+1-dimensional Euclidean space $R^{d+1}$. At each point of
$\cM$ there is a vector \be |n\rangle=\begin{pmatrix}n^1\cr .\cr
..\cr .\cr n^{d+1}\end{pmatrix},\ee which is normal to $\cM$. (Here
$n^i$ are the Cartesian components of $\v n$.) In the following
discussion, Latin indices (refer to the embedding space) range
from 1 to d+1 and Greek indices (refer to the embedded space)
range from 1 to d.

Given the normal vector $\v n$, it is possible to construct the
following $(d+1)\times (d+1)$ matrix \be H\equiv |n\rangle \langle
n|.\label{Hnn}\ee In the above, $\langle n|$, the dual of $|n
\rangle$, is the (d+1)-component row vector \be \langle
n|=\begin{pmatrix}n^1,\ ...,\ n^{d+1}\end{pmatrix}.\ee Obviously
\be \langle n|n \rangle=1.\label{p1}\ee In both \Eq{Hnn}  and
\Eq{p1}, matrix product is assumed. Because $|n \rangle$ and
$\langle n|$ depend on the location in $\cM$, so does $H$. We will
take such matrix as our ``Hamiltonian''.

The matrix defined in \Eq{Hnn} has one eigenvalue $\lambda=1$ and
eigenvalues $\lambda=0$ with d-fold degeneracy. The $\lambda=1$
eigenvector is $\v n$ and the $\lambda=0$ eigenvectors are
orthogonal to $\v n$ and hence are tangent vectors in $\cM$. Since
the $\lambda=0$ eigenvalue is d-fold degenerate, the associated
Berry connection is a $d\times d$ matrix.

Let $|\psi_1 \rangle, ...,|\psi_d \rangle$ be the $d$ orthonormal
eigenvectors of $H$ satisfying \be H|\psi_\mu \rangle=0,
~\mu=1,...,d\label{eig}\ee Obviously, $|\psi_\mu \rangle$ defines
a local orthonormal basis in the tangent space of $\cM$. As the
result, an infinitesimal displacement vector is given by \be d\v
x= dx^{\mu}|\psi_\mu \rangle,\ee which implies \be ds^2= g^\mu_\nu
dx_\mu dx^\nu,\ee where \be g^\mu_\nu=\langle\psi^\mu|\psi_\nu
\rangle.\ee

The non-Abelian Berry's connection is defined as \be
(A_\mu)^\alpha_\beta=\langle\psi^\alpha|\p_\mu|\psi_\beta
\rangle.\ee It can be shown straightforwardly that\be
(A_{\mu})^{\alpha}_{\beta}=\Gamma^{\alpha}_{\beta\mu},\label{rm}\ee
where $\Gamma^{\alpha}_{\beta\mu}$ is the Christoffel connection.
The Riemann-Christoffel curvature tensor is given by \be
R^{\beta}_{\nu\rho\s}\equiv\Gamma^{\beta}_{\nu\s,\rho}-\Gamma^{\beta}_{\nu\rho,\s}+\Gamma^{\beta}_{\alpha\rho}
\Gamma^{\alpha}_{\nu\s}-\Gamma^{\beta}_{\alpha\s}\Gamma^{\alpha}_{\nu\rho}.\label{rie}\ee
Substituting \Eq{rm} into \Eq{rie}, we obtain \be
R^{\beta}_{\nu\rho\s}&&=\{\p_{\rho}A_{\s}-\p_{\s}A_{\rho}+[
A_{\rho},A_{\s}]\}^{\beta}_{\nu}\nn
&&=-(F_{\rho\s})^{\beta}_{\nu}.\label{rie1}\ee
\\

\end{example}

\section{Chern Number and Chern-Simons Invariant\label{sec: Chern-Simons}}

Given a fiber $F$, a structure group $G$ and a (closed) base space
$\cM$, we may construct fiber bundles. A fiber bundle is locally
the direct product of the fiber and the base space. However
globally speaking, after non-trivial twisting of the fibers as the
base space is traversed, a topologically non-trivial bundle can be
produced. The question is by allowing all possible twistings how
many distinct topological types of fiber bundles one can
construct. In mathematics, the characteristic classes are
invariants measuring the topological \textquotedblleft
non-triviality \textquotedblright\ of a fiber bundle. Among them,
the \emph{Chern classes} and the \emph{Chern-Simons classes} are
of particular interests.

\subsection{Chern Number\label{subsec: Chern number}}

If the bases space is a closed 2-dimensional base space $\cM$, the
following integral, the \emph{Chern number} \cite{Chern},
\begin{equation}
C=\int_{\cM}ch_{1}(\mathcal{F})=\frac{1}{8\pi}\int_{\cM}d^{2}x\,\varepsilon
^{\mu\nu}\,Tr[F_{\mu\nu}], \label{cn}\end{equation} is a
topological invariant. If $C$ is nonzero the fiber bundle must
have nontrivial topology. For the $U(1)$ bundle, $C$ is quantized
as $C=n/2$, where $n$ is an integer.

\begin{example}
\label{ex: monopole 2}\noindent Following example \ref{ex:
monopole 1}, consider $H=x\sigma_{x}+y\sigma_{y}+z\sigma_{z}$.
However this time we restrict $(x,\,y,\,z)$ to lie on the surface
of the unit sphere in three dimension ($S^{2}$). For each point on
$S^2$ we use the collection of all normalized eigenvectors of $H$
with eigenvalue $+1$ as the fiber to construct a $U(1)$ fiber
bundle over $S^2$. [The structure group is $U(1)$ because the
normalized eigenvectors are related to one another by $U(1)$
transformations.] The connection of such $U(1)$ bundle is exactly
the Berry connection discussed earlier. If we choose the polar
angle $(\theta,\,\phi)$ to parameterize $S^{2}$, then the
eigenvector correspond to $+1$ eigenvalue is given by
\be
|\psi_+^S\rangle=\begin{pmatrix}e^{-i\phi}\cos(\theta/2)\cr\sin(\theta/2)\end{pmatrix},
\quad
|\psi_+^N\rangle=\begin{pmatrix}\cos(\theta/2)\cr e^{i\phi}\sin(\theta/2)\end{pmatrix},
\ee
where $|\psi_+^S\rangle$ is well-defined in the south hemisphere while $|\psi_+^N\rangle$
is in the north hemisphere and these two differ by a gauge transformation.

The Berry connection is given by
\be
&A_{\theta}^{S/N} =-i\langle\psi_{+}|\partial_{\theta}\psi_{+}\rangle=0,\nonumber\\
&A_{\phi}^S =-i\langle\psi_{+}|\partial_{\phi}\psi_{+}\rangle=-\cos^{2}(\theta/2),
\quad
A_{\phi}^N = \sin^2(\theta/2).
\ee
\label{eq: monopole A+ A-}
As a result, $F_{\theta\phi}=\frac{1}{2}\sin\theta$, and
\begin{align}
C  &  =\frac{1}{8\pi}\int d^{2}x\,\varepsilon^{\mu\nu}F_{\mu\nu},\nonumber\\
&  =\frac{1}{8\pi}\int_{0}^{\pi}d\theta\int_{0}^{2\pi}d\phi\,\sin\theta%
=\frac{1}{2},
\end{align}
since the Chern number is non-zero the $U(1)$ fiber bundle we
constructed is non-trivial.
\end{example}

As we discussed earlier, the Chern number of $U(1)$ bundles must
equal to $n/2$. In the above example we encountered a case where
$n=1$. Can we construct a $2\times 2$ Hamiltonian over $S^2$ such
that the Chern number is greater than $1/2$? The answer is
affirmative due to the following theorem proven by Hsiang and
Lee \cite{HsiangLee}.

Consider the following $2\times 2$ Hamiltonian defined over base
space $S^2$: \be
H=\hat{h}(\theta,\phi)\cdot\vec{\sigma},\label{gh}\ee where
$\vec{\sigma}$ are the Pauli matrices and $\hat{h}(\theta,\phi)$
is a three-component unit vector field (which maps
$S^{2}\rightarrow S^{2}$). This Hamiltonian has eigenvalues $\pm1$
and the corresponding eigenstates $|\psi_{\pm
}(\theta,\phi)\rangle$. Now consider the Berry connection \be
&&A_{\theta}=-i\langle\psi_{+}(\theta,\phi)|\partial_{\theta}|\psi_{+}(\theta,\phi)\rangle,\nn
&&A_{\phi}=-i\langle\psi_{+}(\theta,\phi)|\partial_{\phi}|\psi_{+}(\theta,\phi)\rangle,\ee
associated with, say, the $+$ eigenstate.
\begin{theorem}
The Berry curvature, $F_{\theta\phi}$, associated with the above
connection satisfies
\begin{equation}
F_{\mu\nu}=\frac{1}{2}\hat{h}\cdot(\partial_{\nu}\hat{h}\times\partial_{\mu
}\hat{h}). \label{hl}
\end{equation}
\noindent
\end{theorem}

\begin{proof}%
\begin{align}
F_{\mu\nu}  &  =-i[\partial_{\nu}\langle\psi_{+}|\partial_{\mu}\psi_{+}%
\rangle-(\mu\leftrightarrow\nu)]\nonumber\\
&  =-i[\langle\partial_{\nu}\psi_{+}|\partial_{\mu}\psi_{+}\rangle
-(\mu\leftrightarrow\nu)]\nonumber\\
&  =-i\sum_{n=\pm}[\langle\partial_{\nu}\psi_{+}|\psi_{n}\rangle\langle
\psi_{n}|\partial_{\mu}\psi_{+}\rangle-(\mu\leftrightarrow\nu)].
\end{align}
Since $\langle\psi_{+}|\psi_{+}\rangle=1$, $\langle\partial_{\nu}\psi_{+}%
|\psi_{+}\rangle+\langle\psi_{+}|\partial_{\nu}\psi_{+}\rangle=0$ and
$\langle\partial_{\nu}\psi_{+}|\psi_{+}\rangle$ is purely imaginary. Then
$\langle\partial_{\nu}\psi_{+}|\psi_{+}\rangle\langle\psi_{+}|\partial_{\mu
}\psi_{+}\rangle=\langle\partial_{\mu}\psi_{+}|\psi_{+}\rangle\langle\psi
_{+}|\partial_{\nu}\psi_{+}\rangle$ is real and thus
\begin{equation}
F_{\mu\nu}=-i[\langle\partial_{\nu}\psi_{+}|\psi_{-}\rangle\langle\psi
_{-}|\partial_{\mu}\psi_{+}\rangle-(\mu\leftrightarrow\nu)].
\end{equation}
Form the first order perturbation theory,
\begin{equation}
\langle\psi_{-}|\partial_{\mu}\psi_{+}\rangle=\frac{\langle\psi_{-}%
|\partial_{\mu}\hat{h}\cdot\vec{\sigma}|\psi_{+}\rangle}{E_{+}-E_{-}}%
=\frac{\langle\psi_{-}|\partial_{\mu}\hat{h}\cdot\vec{\sigma}|\psi_{+}\rangle
}{2}.
\end{equation}
Therefore,
\begin{align}
F_{\mu\nu}  &  =-\frac{i}{4}[\langle\psi_{+}|\partial_{\nu}\hat{h}\cdot
\vec{\sigma}|\psi_{-}\rangle\langle\psi_{-}|\partial_{\mu}\hat{h}\cdot
\vec{\sigma}|\psi_{+}\rangle-(\mu\leftrightarrow\nu)]\nonumber\\
&  =-\frac{i}{4}\sum_{n=\pm}[\langle\psi_{+}|\partial_{\nu}\hat{h}\cdot
\vec{\sigma}|\psi_{n}\rangle\langle\psi_{n}|\partial_{\mu}\hat{h}\cdot
\vec{\sigma}|\psi_{+}\rangle-(\mu\leftrightarrow\nu)]\nonumber\\
&  =-\frac{i}{4}\langle\psi_{+}|[\partial_{\nu}\hat{h}\cdot\vec{\sigma
},\partial_{\mu}\hat{h}\cdot\vec{\sigma}]|\psi_{+}\rangle\nonumber\\
&  =-\frac{i}{4}\partial_{\nu}h_{\alpha}\partial_{\mu}h_{\beta}\langle\psi
_{+}|[\sigma_{\alpha},\sigma_{\beta}]|\psi_{+}\rangle\nonumber\\
&  =\frac{1}{2}\varepsilon_{\alpha\beta\gamma}\partial_{\nu}h_{\alpha}%
\partial_{\mu}h_{\beta}\langle\psi_{+}|\sigma_{\gamma}|\psi_{+}\rangle
,\ \ \alpha,\beta,\gamma\in{x,y,z}.
\end{align}
Consider $|\psi_{+}\rangle=h_{\gamma}\sigma_{\gamma}|\psi_{+}\rangle$ and
${\sigma_{\gamma}}^{2}=1$; thus, we have
\begin{equation}
F_{\mu\nu}=\frac{1}{2}\varepsilon_{\alpha\beta\gamma}\partial_{\nu}h_{\alpha
}\partial_{\mu}h_{\beta}h_{\gamma}=\frac{1}{2}\hat{h}\cdot(\partial_{\nu}%
\hat{h}\times\partial_{\mu}\hat{h}).
\end{equation}

\end{proof}

This theorem tells us that in order to construct a Hamiltonian whose
eigenvector bundle exhibits $C>1/2$ we need to choose $\hat{h}$ so
that \be
\cP\equiv\frac{1}{8\pi}\int_{S^2} d^{2}x\,\varepsilon^{\mu\nu}\hat{h}\cdot(\partial_{\nu}%
\hat{h}\times\partial_{\mu}\hat{h})\ee is greater than $1$. The
$S^2\ra S^2$ identity map \be \hat{h}(\hat{r})=\hat{r}\ee gives
$\cP=1$ and the corresponding Hamiltonian [\Eq{gh}] reduces to
that discussed in Example 3. The quantity $\cP$, the Pontrjagin
index, is a homotopy class of the $S^2\ra S^2$ mapping. It
measures the number of times the image covers the target space. A
map that gives $\cP>1$ can be explicitly constructed as follows:
\begin{equation}
\hat{h}(\theta,\phi)=(\sin\alpha(\theta,\phi)\cos\beta(\theta,\phi),\,\sin\alpha(\theta,\phi)\sin\beta(\theta,\phi),\,
\cos\alpha(\theta,\phi)) \label{pont1}\end{equation} with
\begin{align}
\alpha (\theta,\phi) &  =2\cot^{-1}[\cot^{n}(\theta/2)],\nonumber\\
\beta(\theta,\phi) &  =n\phi. \label{pont2}\end{align} In the
above equation, $n$ is an integer that counts the number of times
the image covers the target space. With this $\hat{h}$, the
Pontrjagin index is $\cP=n$. Consequently  by \Eq{cn} and \Eq{hl}
the Chern number is given by \be C=\frac{1}{16\pi}\int_{S^2}
d^{2}x\,\varepsilon^{\mu\nu}\hat{h}\cdot(\partial_{\nu}
\hat{h}\times\partial_{\mu}\hat{h})={\cP\over 2}={n\over 2}.\ee
For $n>1$ this gives the desired nontrivial bundle. Furthermore
because the Pontrjagin index is a topological invariant, all maps
having the same homotopy as that in \Eq{pont1} and \Eq{pont2} lead
to the same Chern number \cite{HsiangLee}.

Another way to modify Chern number is to keep $\hat{h}$ the
identity map but change the dimension of the $\sigma$'s. For
example, if we replace the $2\times 2$ Pauli matrices by the
$3\times 3$ matrices associated with the spin 1 representation of
$SU(2)$:
\begin{equation}
\Sigma_{x}\ra\frac{1}{\sqrt{2}}\left(
\begin{array}
[c]{ccc}%
0 & 1 & 0\\
1 & 0 & 1\\
0 & 1 & 0
\end{array}
\right)  ,\quad \Sigma_{y}\ra\frac{1}{\sqrt{2}}\left(
\begin{array}
[c]{ccc}%
0 & -i & 0\\
-i & 0 & -i\\
0 & i & 0
\end{array}
\right)  ,\quad \Sigma_{z}\ra\frac{1}{\sqrt{2}}\left(
\begin{array}
[c]{ccc}%
1 & 0 & 0\\
0 & 0 & 0\\
0 & 0 & -1
\end{array}
\right)  ,
\end{equation}
then the Chern number $C$ associated with the $U(1)$ bundle whose
fiber is the $\lambda=+1$ eigenvectors of the resulting $3\times
3$ Hamiltonian is 1.

\subsection{Chern-Simons Invariant}

The Chern number [\Eq{cn}] records a specific topological property
of fiber bundles. In the case of $U(1)$ bundle it counts the
number of monopoles (in the connection) enclosed by the (closed)
2D base space. If we have a $U(1)$ bundle over a three dimensional
base space whose connection has no monopole, the Chern number will
vanish for all closed 2D surfaces. Under that condition the dual
of the $F_{\mu\nu}$, i.e.,
$B^\mu=\epsilon^{\mu\nu\lambda}F_{\nu\lambda}$ must form closed
loops in 3D. Now we encounter the next level of topological
intricacy, namely, different $B^\mu$ loops can \emph{link} with
one another!

 This new level of topological non-triviality is
reflected by the \emph{Chern-Simons invariant} \cite{Chern-Simons}:
\begin{equation}
CS=\int_{\cM}Q_{3}(\mathcal{A},\mathcal{F})=\frac{1}{8\pi}\int_{\cM}%
d^{3}x\,\epsilon^{\mu\nu\lambda}A_{\mu}F_{\nu\lambda}.
\label{eq: Chern-Simons}%
\end{equation}
For non-Abelian connection (gauge field) the Chern-Simons
invariant reads as
\begin{equation}
CS=\frac{1}{8\pi}\int_{\cM}d^{3}x\,\varepsilon^{\mu\nu\lambda}Tr[A_{\mu}%
F_{\nu\lambda}].
\end{equation}
It turns out that the Chern-Simons invariant is not quantized for
the U(1) connection but is quantized $(=m^{2}/8\pi)$ for
non-Abelian connections ($m$ is an integer).  It is important to
note that $CS$ is gauge-invariant for $U(1)$ group if $\cM$ has no
boundary. This
follows from the identity $\partial_{\lambda}F_{\mu\nu}+\partial_{\nu}%
F_{\lambda\mu}+\partial_{\mu}F_{\nu\lambda}=0$ and the antisymmetry of
$\varepsilon^{\mu\nu\lambda}$.

In \cite{HsiangLee}, Hsiang and Lee asked the interesting question:
``Is it possible to construct simple $2\times 2$ Hamiltonian over
3-dimensional base space, so that the Chern-Simons invariant is
non-zero?'' In the following we reproduce their results.

\begin{example}
\label{ex: CS 1} Consider the three-sphere $S^3$ parameterized by
\begin{align}
S^{3}  &  =\left\{
(\cos(t/2)\cos\alpha,\,\cos(t/2)\sin\alpha,\,\sin
(t/2)\cos\beta,\,\sin(t/2)\sin\beta)\in R^4;\right. \nonumber\\
&  \left.  0\leq t\leq\pi,\,0\leq\alpha,\beta<2\pi\right\}  .
\label{eq: S3 coordinates}%
\end{align}
Now consider the Hamiltonian \be
H=\hat{h}(t,\alpha,\beta)\cdot\vec{\sigma}\label{s3h} \label{eq: h
dot sigma} \ee with $\hat{h}:S^{3}\rightarrow S^{2}$ being the
nontrivial \emph{Hopf map} specified by the following unit-vector
function:%
\begin{equation}
\hat{h}(t,\alpha,\beta)=(\sin\theta\cos\phi,\,\sin\theta\sin\phi,\,\cos
\theta), \label{eq: Hopf map}%
\end{equation}
where
\begin{align}
&  \theta(t,\alpha,\beta)=2\cot^{-1}[\cot^{m}(t/2)], \nonumber\\
&  \phi((t,\alpha,\beta)=m(\alpha-\beta). \label{eq: Hopf map 2}%
\end{align}
Here $m$ is an integer. With this particular Hopf map, the
Chern-Simons invariant in \Eq{eq: Chern-Simons} computed from the
Berry connection (of the $+1$ eigenstate) is $CS=m^{2}/8\pi$
\cite{HsiangLee}. We will concentrate on this case hereafter.

The dual of the Berry curvature (associated with the $\lambda=+1$)
is given by: \be
B^\mu\equiv\varepsilon^{\mu\nu\lambda}F_{\nu\lambda}=\frac{1}{2}\varepsilon
^{\mu\nu\lambda}\hat{h}\cdot
(\partial_{\lambda}\hat{h}\times\partial_{\nu}\hat{h}). \ee Due to
\Eq{eq: Hopf map 2}, $\partial_\alpha \hat{h}=-\partial_\beta
\hat{h}$. As a result $B^t=0$ and $B^\alpha=B^\beta$.
Consequently, the dual Berry curvature $B^\mu$ is tangent to the
one dimensional lines characterized by $t=constant$ and
$\alpha-\beta=constant$.

\begin{figure}
\begin{picture}(460,220)(0,0)


 \put(10,25){\scalebox{0.9}{\includegraphics{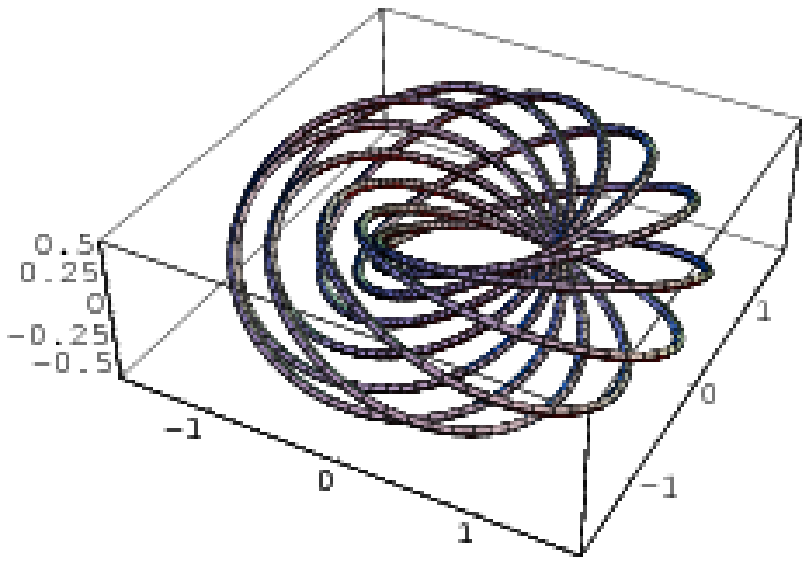}}}
 \put(260,5){\scalebox{0.9}{\includegraphics{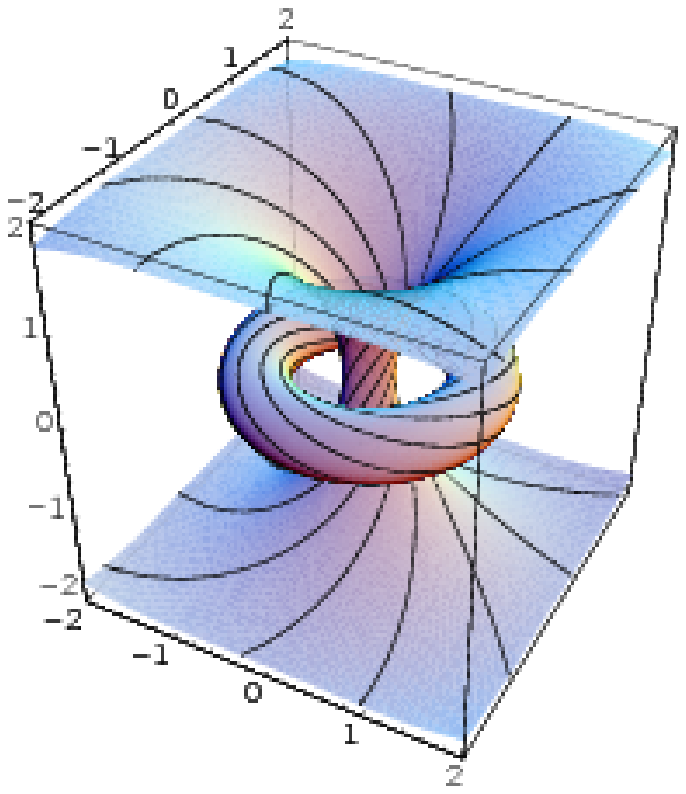}}}

 \put(10,200){(a)}
 \put(260,200){(b)}

\end{picture}
\caption{ The cluster composed of magnetic flux loops for
$\alpha-\beta=0,\ 0.2\pi,\ 0.4\pi,\ \cdots,\ 1.8\pi$ is shown in
(a) with $t=0.3\pi$. The magnetic loops are drawn with finite
thickness in order to illustrate the mutual linking in
perspective. The loops with the same $t$ sweep out the
surface of a torus. The tori with smaller $t$'s are inclosed by
the tori with bigger $t$'s. Thus, every pair of loops is mutually
linked. In particular, we draw the tori for $t=0.2\pi$ and
$t=0.7\pi$ in (b). } \label{fig1}
\end{figure}

To visualize the flux lines associated with the connection, we
employ the stereographic projection w.r.t the north pole ($t=\pi$,
$\beta=\pi/2$) of $S^3$ and map the base space to $
R^3\cup\{\infty\}$ by:
\begin{equation}
x=\frac{\cos(t/2)\cos\alpha}{1-\sin(t/2)\sin\beta}, \quad
y=\frac{\cos(t/2)\sin\alpha}{1-\sin(t/2)\sin\beta}, \quad
z=\frac{\sin(t/2)\cos\beta}{1-\sin(t/2)\sin\beta}. \label{proj}
\end{equation}
Since the stereographic projection is a conformal map, the dual
Berry curvature $B^\mu$ is still tangent to the lines specified by
constant $t$ and $\alpha-\beta$ in $(x,y,z)$-coordinates. In
Fig.~\ref{fig1}, we show the loops formed by $B^\mu$ lines with
different values of $t$ and $\alpha-\beta$. The manifold of
$B^\mu$ loops with the same $t$ form the surface of a torus. All
$B^\mu$ loops on that surface mutually link with one another! In
addition, a torus characterized by $t=t_1$ is inclosed by another
torus $t=t_2>t_1$ such that their $B^\mu$ loops also mutually link
with one another. In sort the loops of the Berry curvature for
the Hamiltonian specified by \Eq{s3h}, \Eq{eq: Hopf map}, \Eq{eq:
Hopf map 2} and \Eq{proj} have the facinating property that every
loop link with all other loops!


\end{example}

\section{The Schr\"{o}dinger Equation\label{sec: Shrodinger}}

Equipped with the above preliminaries, we are ready to study the
Schr\"{o}dinger equation on $S^{3}$ in the presence of a magnetic
field given by the dual Berry curvature $B^\mu$ of \Eq{s3h},
\Eq{eq: Hopf map} and \Eq{eq: Hopf map 2} . In other words, the
vector potential $A_\mu$ that enters the Schr\"{o}dinger equation
is the Berry connection associated with the $+1$ vectors of
\Eq{s3h}, \Eq{eq: Hopf map} and \Eq{eq: Hopf map 2}.

\subsection{The Metric and Laplace Operator}

Again, consider $S^{3}$ as being \ embedded in $R^{4}$ and
parameterized by \Eq{eq: S3 coordinates}. The metric and Laplace
operator on $S^{3}$ are determined from those in $R^{4}$. With the
coordinates in $R^{4}$ given by
\Eq{eq: S3 coordinates} and the metric $ds^{2}=\sum_{\alpha=1}%
^{4}dy_{\alpha}^{2}$, we deduce the metric on $S^{3}$ as%
\begin{equation}
ds^{2}=\frac{1}{4}dt^{2}+\cos^{2}(t/2)d\alpha^{2}+\sin^{2}(t/2)d\beta
^{2}\equiv g_{\mu\nu}dx^{\mu}dx^{\nu};
\end{equation}
i.e.%
\begin{equation}
g_{\mu\nu}=\left(
\begin{array}
[c]{ccc}%
1/4 & 0 & 0\\
0 & \cos^{2}(t/2) & 0\\
0 & 0 & \sin^{2}(t/2)
\end{array}
\right)  ,\quad g^{\mu\nu}=\left(
\begin{array}
[c]{ccc}%
4 & 0 & 0\\
0 & 1/\cos^{2}(t/2) & 0\\
0 & 0 & 1/\sin^{2}(t/2)
\end{array}
\right)  . \label{eq: metric}%
\end{equation}

The Laplace operator on $S^{3}$ is deduced from Laplace operator in $R^{4}$, i.e.
$\sum_{\alpha=1}^{4} \partial^{2}/\partial y_{\alpha}^{2}$. The result
is%
\begin{equation}
g^{\mu\nu}[\partial_{\mu}\partial_{\nu}+(\partial_{\mu}\ln v)\partial_{\nu
}]=4\partial_{t}^{2}+4%
\genfrac{(}{)}{}{}{\cos t}{\sin t}%
\partial_{t}+\frac{1}{\cos^{2}(t/2)}\partial_{\alpha}^{2}+\frac{1}{\sin
^{2}(t/2)}\partial_{\beta}^{2},
\end{equation}
where
$v=(\sin t)/4$
is the Jacobian of the coordinate transformation in \Eq{eq: S3
coordinates}.

The Schr\"{o}dinger equation in the presence of the magnetic field
can be set up by replacing $\partial_{\mu
}\rightarrow\partial_{\mu}-iA_{\mu}^{b}$ in the Laplace operator.
Thus we
obtain%
\begin{equation}
-\frac{1}{2M}g^{\mu\nu}\left[
(\partial_{\mu}-iA_{\mu}^{b})(\partial_{\nu
}-iA_{\nu}^{b})+(\partial_{\mu}\ln
v)(\partial_{\nu}-iA_{\nu}^{b})\right]
\Psi=E\Psi, \label{eq: Schrodinger}%
\end{equation}
where $A_{\mu}^{b}$ is the Berry connection associated with the
$+1$ eigenstate $|Z(t,\alpha,\beta)\rangle$ of \Eq{s3h}, \Eq{eq:
Hopf map} and \Eq{eq: Hopf map 2} via
\begin{equation}
A_{\mu}^{b}=-i\langle Z(t,\alpha,\beta|\partial_{\mu}Z(t,\alpha,\beta)\rangle,
\label{eq: Berry connection}%
\end{equation}

\subsection{Computing Berry Connection}

To compute $A_{\mu}^{b}$, we choose the following gauge%
\begin{equation}
|Z(t,\alpha,\beta)\rangle=\binom{e^{im\alpha}\cos[\theta(t)/2]}{e^{im\beta
}\sin[\theta(t)/2]}, \label{eq: spinor Z}%
\end{equation}
where $\theta(t)$ is given by \Eq{eq: Hopf map 2}. It is easy to
prove that the above
$|Z(t,\alpha,\beta)\rangle$ satisfies%
\begin{equation}
H(t,\alpha,\beta)|Z(t,\alpha,\beta)\rangle=(+1)|Z(t,\alpha,\beta)\rangle
\end{equation}
with $H(t,\alpha,\beta)$ given by \Eq{eq: h dot sigma}.

A simple calculation using \Eq{eq: Berry connection} and
\Eq{eq: h dot sigma} yields %
\begin{equation}
A_{t}^{b}=0,\quad A_{\alpha}^{b}=m\cos^{2}(\theta/2),\quad A_{\beta}^{b}%
=m\sin^{2}(\theta/2), \label{eq: Berry A}%
\end{equation}
where $\theta(t)$ is given by \Eq{eq: Hopf map 2}.

Substituting \Eq{eq: metric} and \Eq{eq: Berry A} into
\Eq{eq: Schrodinger}, we obtain the Schr\"{o}dinger equation%

\begin{gather}
-\frac{2}{M}\left\{  \partial_{t}^{2}+\left(  \frac{\cos t}{\sin t}\right)
\partial_{t}+\frac{1}{4\cos^{2}(t/2)}[\partial_{\alpha}-i\,m\cos^{2}%
(\theta/2)]^{2}\right. \nonumber\\
\left.  +\frac{1}{4\sin^{2}(t/2)}[\partial_{\beta}-i\,m\sin^{2}(\theta
/2)]^{2}\right\}  \Psi=E\Psi. \label{eq: Shrodinger 2}%
\end{gather}

\section{Solutions without Magnetic Field --- $SO(4)$ Spherical
Harmonics\label{sec: spherical harmonics}}

In the absence of magnetic field, the Hamiltonian of \Eq{eq:
Shrodinger 2} (with $m=0$) is invariant under $SO(4)$
transformation. Consider the six generators of $SO(4)$:
\be
  \hat{L}_{ij}=r_{i}p_{j}-r_{j}p_{i},
   \ee where $\vec{r}=(y_{1},y_{2},y_{3},y_{4})$ and $\vec{p}=-i(\partial_{y_{1}}%
,\partial_{y_{2}},\partial_{y_{3}},\partial_{y_{4}})$.
   Each of $\hat{L}_{ij}$ corresponds to the rotation on the
plane spanned by $y_{i}$ and $y_{j}$. Regrouping these operators
as
\begin{align}
&  \vec{L}=(L_{23},L_{31},L_{12}),\nonumber\\
&  \vec{M}=(L_{14},L_{24},L_{34}),
\end{align}
we obtain the following commutation relations:
\begin{align}
&  [L_{i},L_{j}]=i\epsilon_{ijk}L_{k}\nonumber\\
&  [M_{i},L_{j}]=i\epsilon_{ijk}M_{k}\nonumber\\
&  [M_{i},M_{j}]=i\epsilon_{ijk}L_{k}. \label{commt}
\end{align}

In the $(t,\alpha,\beta)$-coordinates, these operators read as
\begin{align}
&  L_{1}=L_{23}=\frac{1}{i}\left\{  2\sin\alpha\cos\beta\,\partial_{t}%
-\tan\frac{t}{2}\cos\alpha\cos\beta\,\partial_{\alpha}-\cot\frac{t}{2}%
\sin\alpha\sin\beta\,\partial_{\beta}\right\}  ,\nonumber\\
&  L_{2}=L_{31}=\frac{1}{i}\left\{  -2\cos\alpha\cos\beta\,\partial_{t}%
-\tan\frac{t}{2}\sin\alpha\cos\beta\,\partial_{\alpha}+\cot\frac{t}{2}%
\cos\alpha\sin\beta\,\partial_{\beta}\right\}  ,\nonumber\\
&  L_{3}=L_{12}=\frac{1}{i}\,\partial_{\alpha},
\end{align}
and
\begin{align}
&  M_{1}=L_{14}=\frac{1}{i}\left\{  2\cos\alpha\sin\beta\,\partial_{t}%
+\tan\frac{t}{2}\sin\alpha\sin\beta\,\partial_{\alpha}+\cot\frac{t}{2}%
\cos\alpha\cos\beta\,\partial_{\beta}\right\}  ,\nonumber\\
&  M_{2}=L_{24}=\frac{1}{i}\left\{  2\sin\alpha\sin\beta\,\partial_{t}%
-\tan\frac{t}{2}\cos\alpha\sin\beta\,\partial_{\alpha}+\cot\frac{t}{2}%
\sin\alpha\cos\beta\,\partial_{\beta}\right\}  ,\nonumber\\
&  M_{3}=L_{34}=\frac{1}{i}\,\partial_{\beta}.
\end{align}
A straightforward calculation using the above results gives
\begin{align}
 & \vec{L}^{2}+\vec{M}^{2}
  =L_{1}^{2}+L_{2}^{2}+L_{3}^{2}+M_{1}^{2}+M_{2}^{2}+M_{3}^{2}\nonumber\\
=&-\left\{  4\partial_{t}^{2}+4\left(  \frac{\cos t}{\sin t}\right)
\partial_{t}+\frac{1}{\cos^{2}(t/2)}\,\partial_{\alpha}^{2}+\frac{1}{\sin
^{2}(t/2)}\,\partial_{\beta}^{2}\right\},\label{mres}
\end{align}
and
\begin{equation}
\vec{L}\cdot\vec{M}=\vec{M}\cdot\vec{L}=0. \label{eq: L dot M}%
\end{equation}
\Eq{mres} implies that
\begin{equation}
-\nabla_{spherial}^{2}=\vec{L}^{2}+\vec{M}^{2}.
\end{equation}

By introducing $\vec{I}=(\vec{L}+\vec{M})/2$ and $\vec{K}=
(\vec{L}-\vec{M})/2$, \Eq{commt} becomes
\begin{align}
&  [I_{i},K_{j}]=0\nonumber\\
&  [I_{i},I_{j}]=i\,\epsilon_{ijk}I_{k}\nonumber\\
&  [K_{i},K_{j}]=i\,\epsilon_{ijk}K_{k}.
\end{align}
Consequently the Lie algebra $so(4)$ is equivalent to $so(2)\oplus
so(2)$, of which the representation states are $|j_{I}m_{I}\rangle
\otimes|j_{K}m_{K}\rangle$. Furthermore due to \Eq{eq: L dot M}, $\vec{I}%
^{2}=\vec{K}^{2}$, thus $j_{I}=j_{K}\equiv j$. Therefore the
representation states for $SO(4)$ are
\begin{align}
&  |jm_{I}\rangle\otimes|jm_{K}\rangle\equiv|jm_{I}m_{K}\rangle,\nonumber\\
&  m_{i},\ m_{j}=-j,\,-j+1,\cdots,j-1,\,j\nonumber\\
&  j=0,\,1/2,\,1,\,3/2,\cdots
\end{align}
These states satisfy the following equations
\begin{align}
&  \vec{I}^{2}|jm_{I}m_{K}\rangle=\vec{K}^{2}\,|jm_{I}m_{K}\rangle
=j(j+1)\,|jm_{I}m_{K}\rangle,\nonumber\\
&  I_{3}\,|jm_{I}m_{K}\rangle=m_{I}\,|jm_{I}m_{K}\rangle,\nonumber\\
&  K_{3}\,|jm_{I}m_{K}\rangle=m_{K}\,|jm_{I}m_{K}\rangle.
\end{align}

Since $-\nabla_{spherical}^{2}=\vec{L}^{2}+\vec{M}^{2}=2(\vec{I}^{2}%
+\vec{K}^{2})$, the states $|jm_{I}m_{K}\rangle$ are also the
eigenstates of the Hamiltonian because \be H=-{1\over 2M}{\nabla_{spherical}^{2}}=\frac{1}{M}(\vec{I}^{2}+\vec{K}%
^{2}),\ee hence \be
H\,|jm_{I}m_{K}\rangle=\frac{2}{M}j(j+1)|jm_{I}m_{K}\rangle.
\label{eq: eigenvalue of H}\ee The wavefunctions
$Y_j^{m_1,m_2}(t,\alpha,\beta)$ (defined below) corresponding to
$|jm_{I}m_{k}\rangle$ are the the 4-dimensional spherical
harmonics. To drive them we first express the generators in
$(t,\alpha,\beta)$-coordinates.

Firstly, $I_{3}=(L_{3}%
+M_{3})/2=(\partial_{\alpha}+\partial_{\beta})/2i$ and $K_{3}%
=(L_{3}-M_{3})/2=(\partial_{\alpha}-\partial_{\beta})/2i$ lead to
$-i\partial_{\alpha}=I_{3}+K_{3}$ and $-i\partial_{\beta
}=I_{3}-K_{3}$; hence
\begin{align}
&  \frac{1}{i}\,\partial_{\alpha}\,|jm_Im_K\rangle=(m_{I}+m_{K}%
)|jm_Im_K\rangle,\nonumber\\
&  \frac{1}{i}\,\partial_{\beta}\,|jm_Im_K\rangle=(m_{I}-m_{K}%
)|jm_Im_K\rangle.
\end{align}
The solution of the above equations is given by the following \be
|jm_Im_K\rangle\equiv Y_j^{m_1,m_2}(t,\alpha,\beta)
=P_{j}^{m_{1},m_{2}}(t)\,e^{im_{1}\alpha}e^{im_{2}\beta},\label{eq:
J-mi-mk}\ee where \be
 m_{1}\equiv m_{I}+m_{K},\ \ m_{2}\equiv m_{I}-m_{K}. %
\ee

Next we define the raising and lowering operators,
\begin{align}
&  L_{\pm}=L_{1}+iL_{2}=\frac{e^{\pm i\alpha}}{i}\left\{  \mp2i\cos
\beta\,\partial_{t} -\tan\frac{t}{2}\cos\beta\,\partial_{\alpha}\pm i\cot
\frac{t}{2}\sin\beta\,\partial_{\beta}\right\}  ,\nonumber\\
&  M_{\pm}=M_{1}+iM_{2}=\frac{e^{\pm i\alpha}}{i}\left\{  2\sin\beta
\,\partial_{t} \mp i\tan\frac{t}{2}\sin\beta\,\partial_{\alpha}+\cot\frac
{t}{2}\cos\beta\,\partial_{\beta}\right\}  ,
\end{align}
and correspondingly,
\begin{align}
&  I_{\pm}=\frac{1}{2}(L_{\pm}+M_{\pm})=\frac{e^{\pm i(\alpha+\beta)}}{2i}
\left\{  \mp2i\,\partial_{t} -\tan\frac{t}{2}\,\partial_{\alpha}+\cot\frac
{t}{2}\,\partial_{\beta}\right\}  ,\nonumber\\
&  K_{\pm}=\frac{1}{2}(L_{\pm}-M_{\pm})=\frac{e^{\pm
i(\alpha-\beta)}}{2i} \left\{  \mp2i\,\partial_{t}
-\tan\frac{t}{2}\,\partial_{\alpha}-\cot\frac
{t}{2}\,\partial_{\beta}\right\}.
\end{align}
These operators satisfy
\begin{align}
I_{\pm}|jm_{I}m_{K}\rangle &  = [(j\mp m_{I})(j\pm m_{I}+1)]^{1/2}%
\;|j,m_{I}\pm1,m_{K}\rangle\nonumber\\
K_{\pm}|jm_{I}m_{K}\rangle &  = [(j\mp m_{K})(j\pm m_{K}+1)]^{1/2}%
\;|j,m_{I},m_{K}\pm1\rangle. \label{dfg}
\end{align}

Now let us consider the highest-weight state with $m_{I}=j$ and $m_{K}=j$
\be
|jjj\rangle=Y_j^{2j,0}(t,\alpha,\beta)=P_j^{2j,0}(t)\,e^{i2j\alpha}\ee
Since this state is annihilated by $I_+$ and $K_+$ we require \be \left(
-2i\,\partial_{t}-\tan\frac{t}{2}\,\partial_{\alpha}\pm\cot\frac
{t}{2}\,\partial_{\beta}\right)  \left[
P_j^{2j,0}(t)e^{i2j\alpha}\right]  =0,\ee or \be
\frac{d\,P_j^{2j,0}(t)}{dt}+j\,\tan\frac{t}{2}\,P_j^{2j,0}(t)=0.\ee
The solution of the above equation is \be
P_j^{2j,0}(t)\propto(\cos\frac{t}{2})^{2j}. \ee Hence the
(normalized)  highest weight state is given by
\begin{equation}
|jjj\rangle=Y_{j}^{2j,0}(t,\alpha,\beta)=\frac{1}{\pi}\sqrt{\frac{1+2j}{2}}\left(
\cos \frac{t}{2}\right)  ^{2j}e^{2ij\alpha}.
\end{equation}

Applying $(I_{-})^{j-m_{I}}(K_{-})^{j-m_{K}}$ on
$Y_{j}^{2j,0}(t,\alpha,\beta)$ by \Eq{dfg}, we generate
$Y_{j}^{m_1,m_2}(t,\alpha,\beta)$. (Here we note $[I_{-},\
K_{-}]=0$ hence the order of the operations does not matter.) In
the following we list the first few
$Y_{j}^{m_1,m_2}(t,\alpha,\beta)$:
\be j=0:\quad &&|0,0,0\rangle=Y_0^{0,0}=
\frac{1}{\sqrt{2}\pi};\ee
\be j=1/2:\quad &&|1/2,1/2,1/2\rangle=Y_{1/2}^{1,0} =\frac{1}{\pi}\cos\frac{t}%
 {2}e^{i\alpha}\nn&&|1/2,1/2,-1/2\rangle=Y_{1/2}^{0,1} =-\frac{1}{\pi}\sin\frac{t}{2}e^{i\beta}\nn
 &&|1/2,-1/2,1/2\rangle=Y_{1/2}^{0,-1}=-\frac{1}{\pi}\sin\frac{t}{2}e^{-i\beta}\nn&&
 |1/2,-1/2,-1/2\rangle=Y_{1/2}^{-1,0}=-\frac{1}{\pi}\cos\frac{t}{2}e^{-i\alpha};\ee
\be j=1:\quad &&|1,1,1\rangle=Y_1^{2,0}=\frac{1}{\pi}\sqrt{\frac{3}{2}}\cos
^{2}(\frac{t}{2})\,e^{2i\alpha}\nn&&|1,1,0\rangle=Y_1^{1,1}=-\frac{\sqrt{3}}{2\pi}\sin
t\,e^{i(\alpha+\beta)}\nn &&|1,1,-1\rangle=Y_1^{0,2}=\frac{1}{\pi}\sqrt{\frac{3}{2}}\sin^{2}(\frac{t}%
{2})\,e^{2i\beta}\nn&&|1,0,1\rangle=Y_1^{1,-1}=-\frac{\sqrt{3}}{2\pi}\sin
t\,e^{i(\alpha-\beta)}\nn&&|1,0,0\rangle=Y_1^{0,0}=-\frac{1}{\pi}\sqrt{\frac{3}{2}}\cos
t\nn&&|1,0,-1\rangle=Y_1^{-1,1}=\frac{\sqrt{3}}{2\pi}\sin t\,e^{-i(\alpha-\beta
)}\nn&&|1,-1,1\rangle=Y_1^{0,-2} =\frac{1}{\pi}\sqrt{\frac{3}{2}}\sin^{2}(\frac{t}%
{2})\,e^{-2i\beta}\nn&&|1,-1,0\rangle=Y_1^{-1,-1}=\frac{\sqrt{3}}{2\pi}\sin
t\,e^{-i(\alpha+\beta )}\nn&&|1,-1,-1\rangle=Y_1^{-2,0}=\frac{1}{\pi}\sqrt{\frac{3}{2}}\cos^{2}(\frac{t}%
{2})\,e^{-2i\alpha}. \ee
The wavefunction
$Y_j^{m_1,m_2}(t,\alpha,\beta)$ is an eigenstate of $H$ with the
eigenvalue $E=2j(j+1)/M$. [See \Eq{eq: eigenvalue of H}.]

\section{Solutions with Magnetic Field of Nonzero Chern-Simons
Invariant\label{sec: nonzero field}}

\subsection{$m=1$}

Now, we return the \Eq{eq: Shrodinger 2} with $m=1$. In this case,
the Hopf map in \Eq{eq: Hopf map 2} reads as
\begin{align}
&  \theta(t,\alpha,\beta)=t,\nonumber\\
&  \phi(t,\alpha,\beta)=\alpha-\beta,
\end{align}
and the Berry connection in \Eq{eq: Berry A} simplifies to
\begin{equation}
A_{t}^{b}=0,\quad A_{\alpha}^{b}=\cos^{2}(t/2),\quad A_{\beta}^{b}%
=\sin^{2}(t/2), \label{eq: Berry B}%
\end{equation}
Substitute the above result into \Eq{eq: Shrodinger 2} we obtain
\begin{gather}
-\frac{2}{M}\left\{  \partial_{t}^{2}+\left(  \frac{\cos t}{\sin t}\right)
\partial_{t}+\frac{1}{4\cos^{2}(t/2)}[\partial_{\alpha}-i\,\cos^{2}%
(t/2)]^{2}\right. \nonumber\\
\left.  +\frac{1}{4\sin^{2}(t/2)}[\partial_{\beta}-i\,\sin^{2}(t/2)]^{2}%
\right\}  \Psi=E\Psi. \label{eq: Schrodinger m=1}%
\end{gather}

In the presence of the vector potential the $SO(4)$ symmetry is
broken. However, \Eq{eq: Schrodinger m=1} is still invariant under
$\alpha\rightarrow \alpha+a$, $\beta\rightarrow\beta+b$. As a
result, we seek for the solution in the form of
\begin{equation}
\Psi(t,\alpha,\beta)=e^{i(m_{1}\alpha+m_{2}\beta)}\,\Theta(t).
\label{eq: factorized form}%
\end{equation}
Substitute the above expression into \Eq{eq: Schrodinger m=1}, we
obtain
\begin{equation}
\lambda\,\Theta=\left\{  -\partial_{t}^{2}-\left(  \frac{\cos
t}{\sin
t}\right)  \partial_{t}+\frac{m_{1}^{2}}{4\cos^{2}(t/2)}+\frac{m_{2}^{2}%
}{4\sin^{2}(t/2)}-\frac{1}{2}(m_{1}+m_{2})+\frac{1}{4}\right\}
\,\Theta, \label{qw}
\end{equation}
where $\lambda=ME/2$.

Let us define $\lambda^{\prime}=\lambda+(m_{1}+m_{2})/2-1/4$, and
reduce \Eq{qw} to
\begin{equation}
\left\{  -\partial_{t}^{2}-\left(  \frac{\cos t}{\sin t}\right)  \partial
_{t}+\frac{m_{1}^{2}}{4\cos^{2}(t/2)}+\frac{m_{2}^{2}}{4\sin^{2}%
(t/2)}\right\}  \,\Theta=\lambda^{\prime}\,\Theta,\label{qw2}
\end{equation}
By substituting \Eq{eq: J-mi-mk} into \Eq{eq: Shrodinger 2}, it is
simple to show that this is exactly the same differential equation
satisfied by $P_{j}^{m_{1},m_{2}}(t)$
in the absence of the
magnetic field ($m=0$), except that $\lambda=ME/2$ is now replaced
by $\lambda^{\prime}$.  Consequently,
$\Theta(t)=P_{j}^{m_{1},m_{2}}(t)$ is till the solution of \Eq{qw2}, and $Y_j^{m_{1},m_{2}}=P_{j}%
^{m_{1},m_{2}}(t)e^{i\,m_{1}\alpha}e^{i\,m_{2}\beta}$ is still the
function of $H$. In other words \be &
H_{m=1}\,Y_j^{m_{1},m_{2}}(t,\alpha,\beta)=E\,
Y_j^{m_{1},m_{2}}(t,\alpha,\beta),\ee where \be
E ={2\over M}\Big[j(j+1)-\frac{m_{1}+m_{2}}{2}+\frac{1}{4}\Big]. \label{eq: eigenvalues for m=1}%
\ee

\subsection{General $m$}

Like the $m=1$ case, \Eq{eq: Shrodinger 2} is still invariant
under $\alpha\rightarrow\alpha+a$, $\beta\rightarrow\beta+b$,
hence the solution has the form of \Eq{eq: factorized form}.
Unfortunately, for $m\ne 0,1$ the equation of $\Theta(t)$ is quite
formidable.

However, as discussed by Hsiang and Lee \cite{HsiangLee}, so long
as the map $\hat{h}(t,\alpha,\beta)$ used in \Eq{s3h} is in the
same homotopy class as the map in \Eq{eq: Hopf map 2}, the
Chern-Simons invariant computed from the Berry connection via
\Eq{eq: Chern-Simons} remain unchanged. This allow us to deform
the map in \Eq{eq: Hopf map 2} to a simpler map while preserving
its homotopy class:
\begin{align}
&  \theta(t,\alpha,\beta)=t,\nonumber\\
&  \phi(t,\alpha,\beta)=m(\alpha-\beta). \label{eq: Hopf map new}%
\end{align}
This deformed Hopf map is continuous and respects the bijection
(one-to-one and onto) from $t$ to $\theta$. As a result, it is in
the same homotopy class as the map in \Eq{eq: Hopf map 2}.

With this map, the magnetic potential computed from the Berry connection is
given by
\begin{equation}
A_{t}^{b}=0,\ \ A_{\alpha}^{b}=m\cos^{2}(t/2),\ \ A_{\beta}^{b}=m\sin
^{2}(t/2).
\end{equation}
Compare this result with \Eq{eq: Berry B} we notice that the new
vector potential is $m$ times that of the $m=1$ case. (Since the
Chern-Simons invariant is quadratic in $A^b_\mu$ this increases
$CS$ by a factor of $m^2$.) When such vector potential is
substituted into \Eq{eq: Shrodinger 2} everything is the same as
$m=1$ case except an overall factor $m$ is involved. Therefore,
the differential equation for $\Theta(t)$ reads
\begin{gather}
\left\{  -\partial_{t}^{2}-\left(  \frac{\cos t}{\sin t}\right)
\partial _{t}+\frac{m_1^2}{4\cos^{2}(t/2)}+\frac{m_2^2}{4\sin^{2}(t/2)}+{m(m-2m_1-2m_2)\over 4}\right\}
\Theta=\lambda\,\Theta.
\end{gather}
Compare the above equation with \Eq{qw2} we note that they only
differ by a redefinition of $\lambda'$. As the result the
eigenstates and eigenvalues of $H(m)$ are
\begin{align}
& H(m)|j\,m_{I}\,m_{K}\rangle=E\,|j\,m_{I}\,m_{K}\rangle,\quad
 E  =2\,\lambda/M,\nonumber\\
& \lambda   =j(j+1)-\frac{m}{2}(m_{1}+m_{2})+\frac{m^{2}}{4}.
\label{eq: eigenvalues for general m}%
\end{align}

From \Eq{eq: eigenvalues for general m}, we see that the effect of
the self-linking magnetic field is to lift the energy level
degeneracy while preserving the eigen-wavefunctions. This is very
similar to the Zeeman effect in elementary quantum mechanics.

We can easily see why this is so by expressing $H$ in terms of $SO(4)$
operators:
\begin{eqnarray}
H  &\equiv& -\frac{2}{M}\left\{  \partial_{t}^{2}+\left(  \frac{\cos t}{\sin
t}\right)  \partial_{t}+\frac{1}{4\cos^{2}(t/2)}[\partial_{\alpha}%
-i\,m\cos^{2}(\theta/2)]^{2}\right. \nonumber \\
& & \mbox{}+ \left.\frac{1}{4\sin^{2}(t/2)}[\partial_{\beta
}-i\,m\sin^{2}(\theta/2)]^{2}\right\} \nonumber\\
&=&-\frac{2}{M}\left\{  \partial_{t}^{2}+\left(  \frac{\cos t}{\sin
t}\right)  \partial_{t}+\frac{\partial_{\alpha}^{2}}{4\cos^{2}(t/2)}%
+\frac{\partial_{\beta}^{2}}{4\sin^{2}(t/2)}-\frac{i}{2}m(\partial_{\alpha
}+\partial_{\beta})-\frac{m^{2}}{4}\right\} \nonumber\\
&=&-\frac{2}{M}\left\{  \frac{\nabla_{spherical}^{2}}{4}-\frac{i}%
{2}m(\partial_{\alpha}+\partial_{\beta})-\frac{m^{2}}{4}\right\} \nonumber\\
&=&\frac{2}{M}\left\{  \frac{1}{2}(\vec{I}^{2}+\vec{K}^{2})-m\,I_{3}%
+\frac{m^{2}}{4}\right\}.
\end{eqnarray}
The above equation makes it clear that $|j\,m_{I}\,m_{K}\rangle$
are eigenstates.

\section{Discussion\label{sec: discussion}}

In this paper, we have solved the Schr\"{o}dinger equation for a
charged particle moving on a {\it 3-sphere} ($S^{3}$) under the
influence of a magnetic field whose flux lines exhibit mutual
linking. (The Chern-Simons invariant associated
with the vector potential is non-zero.)  

The vector potential of the self-linking magnetic field is
obtained from the Berry connection of a simple $2\times 2$
Hamiltonian $H=\hat{h}\cdot\vec{\sigma}$, where $\hat{h}$ is a
three-component unit vector field over $S^3$. It turns out that
when $\hat{h}$ corresponds to a topologically non-trivial mapping
from $S^3$ to $S^2$, the associated Berry's connection exhibits a
non-zero Chern-Simons invariant. 
For simple choice of $\hat{h}$, e.g., the map given in \Eq{eq:
Hopf map new}, the eigenstates are $SO(4)$ spherical harmonics
$Y_{j}^{m_{1},m_{2}}$. However the degeneracy of eigenenergy in
zero field is lifted 
in a manner reminiscent of the Zeeman effect.

This result is very different from the case of a charged particle
moving on a {\it two sphere} under the influence of a Dirac
monopole. In the latter case 
the  $SO(3)$ spherical harmonics $Y_{l,m}$ are not the
eigenfunctions. The real eigen solutions, the monopole harmonics
$Y_{q,l,m}$, have \emph{section} structure. They are the basis for
the irreducible representations of $SU(2)$ which is the universal
covering group of $SO(3)$. This originates from the fact that
there does not exist an everywhere non-singular vector potential for the monopole
field over the entire two-sphere.
On the contrary, there does exist a non-singular vector potential
describing a self-linking magnetic field over the entire $S^3$.


\begin{acknowledgments}
DHL is supported by DOE grant DE-AC03-76SF00098.
\end{acknowledgments}


\begin{thebibliography}{99}

\bibitem {Dirac:1931}P.~A.~M. Dirac, Proc. R. Soc. London,  Ser. A
\textbf{133}, 60 (1931)

\bibitem {Wu-Yang:1975}T.~T.~Wu and C.~N.~Yang, Phys. Rev. D \textbf{12} 3845 (1975)



\bibitem {Nakahara}For a review, see, e.g., ``Geometry, Topology and
Physics" by M.~Nakahara, IOP Publishing (1990), Sec 11.3

\bibitem {Belavin:1975}A.~A.~Belavin, A.~M.~Polyakov, A.~S.~Schwartz and
Y.~S.~Tyupkin, Phys. Lett. \textbf{59B} 85 (1975)


\bibitem {Wu-Yang:1976}T.~T.~Wu and C.~N~Yang, Nucl. Phys. \textbf{B107}
365-380 (1976)

\bibitem {SU(2) monopole} C.N. Yang, J. Math. Phys. {\bf 19}, 320 (1978).

\bibitem{HsiangLee} W-Y Hsiang and D-H Lee, Phys. Rev. A {\bf64},
052101 (2001).

\bibitem {Berry:1984}M.~V.~Berry, Proc. R. Soc. London,
\textbf{A392} 45 (1984)

\bibitem {Simon:1983}B.~Simon, Phys. Rev. Lett. \textbf{51} 2167 (1983)

\bibitem{YM-GR-yang} C. N. Yang, Phys. Rev. Lett. {\bf 33}, 445 (1974).

\bibitem{YM-GR-book} For a review, see ``Quantum Field
Theory'' by L. Ryder, Cambridge University Press (1996), Sec 3.6,
and reference therein

\bibitem {Chern} S.~S.~Chern, Ann. Math. \textbf{47}, 85 (1946)


\bibitem {Chern-Simons}S.~S.~Chern and J.~Simons, Ann. Math.
\textbf{99}, 48 (1974)


\end{thebibliography}
\end{document}